\documentclass[11pt,a4paper]{article}
\usepackage[hyperref]{acl2020}
\usepackage{times}
\usepackage{latexsym}
\usepackage{microtype}
\usepackage{hyperref}
\usepackage{amsmath}
\usepackage{color}
\usepackage{graphicx}
\usepackage{makecell}
\usepackage{amsfonts}
\usepackage{multicol}
\usepackage{xspace}
\usepackage{multirow}

\usepackage{microtype}

\aclfinalcopy 

\providecommand{\keywords}[1]
{
  \small	
  \textbf{\textit{Keywords---}} #1
}
\title{Multilingual Speech Emotion Recognition With Multi-Gating Mechanism and Neural Architecture Search}

\author{
Zihan Wang$^{1}$, Qi Meng$^{1}$,  HaiFeng Lan$^{1}$, XinRui Zhang$^{1}$, KeHao Guo$^{1}$, Akshat Gupta$^{2}$\thanks{*Corresponding author.}~~~~~~~ \\
$^{1}$Columbia University~~~~~~~~~~~~~~\\
$^{2}$JP Morgan AI Research, New York, USA \\
{\tt \small \{zw2782, qm2162, hl3487, xz2976, kg2937\}@columbia.edu} \\
{\tt \small akshat.x.gupta@jpmorgan.com}
}

\date{}

\begin{document}
\maketitle
\begin{abstract}
Speech emotion recognition (SER) classifies audio into emotion categories such as Happy, Angry, Fear, Disgust and Neutral. While Speech Emotion Recognition (SER) is a common application for popular languages, it continues to be a problem for low-resourced languages, i.e., languages with no pretrained speech-to-text recognition models. This paper firstly proposes a language-specific model that extract emotional information from multiple pre-trained speech models, and then designs a multi-domain model that simultaneously performs SER for various languages. Our multi-domain model employs a multi-gating mechanism to generate unique weighted feature combination for each language, and also searches for specific neural network structure for each language through a neural architecture search module. In addition, we introduce a contrastive auxiliary loss to build more separable representations for audio data. Our experiments show that our model raises the state-of-the-art accuracy by 3$\%$ for German and 14.3$\%$ for French.

\end{abstract}

\section{Introduction}

Speech Emotion Recognition is an essential field for current artificial intelligence research. The technique has great potential application in cognitive science, healthcare, and marketing \cite{ser_app}. As mentioned in \cite{multimodal_er}, facial expression contributes for 55$\%$ of the emotional expression, the vocal part contributes for 37$\%$, while the words contribute for 8$\%$. Indeed, the performance of the SER system is excellent with transcripts or ASR \cite{ser_asr}. However, due to the scarcity of emotion-labeled data with transcripts and extensive data \cite{libri} required to train ASR systems in low-resourced languages, learning how to work with audio data alone becomes crucial. 

Two main factors to detect emotion are content information and acoustic information. Previous works generated embeddings for general spoken language tasks in various ways such as phones \cite{allosaurus}\cite{gupta2021acoustics} and Mel-frequency \cite{librosa}. With these pre-trained embeddings, we build a model to gather all these features specifically for SER. Our model uses a CNN + Bi-LSTM + self-attention module to transform sequential features into dense representations. Beyond that, We propose a contrastive auxiliary loss to encourage the model to learn better feature representations. To aid deployment and maintenance in real-world applications, we propose a multi-domain model that trains multiple languages simultaneously. To reduce the effect of negative transfer, we utilize a multi-gating mechanism to form a uniquely weighted combination of embeddings for each language domain. To further increase the flexibility of our model, we introduce a neural architecture search module to design optimal neural network structure for each language automatically. We reach competitive results on English dataset and claim new state-of-the-art accuracy on French and German datasets.

\section{Related Work}

\textbf{Speech Emotion Recognition.} Early work in SER extracts features such as pitch, energy, mel-band energies, and mel-frequency cepstral coefficients (MFCCs) as features, and then use classifiers such as SVMs, LDA, QDA and HMMs to classify audio into a certain emotion class \cite{svm}. In recent years, neural-based models utilize architectures such as CNN and LSTM \cite{cnnlstm}, ResNet-101 \cite{resnet101}, and attention mechanism \cite{att1} in order to achieve better performance. Utilizing the transcripts together with audio file can also enhance SER model performance, since transcripts is a powerful feature in training. Utilizing the transcript can raise the accuracy by around 4\% on IEMOCAP dataset \cite{ser_wordemb}. Applying transfer learning from Automatic Speech Recognition(ASR) to SER is also shown to be a useful method \cite{ser_asr}.

\textbf{Multilingual training.} Other than single-domain SER task, which focuses on one language only, multi-domain SER task, which trains and deploys on multiple languages simultaneously, is also worth studying. Such similar multilingual training is widely applied in Natural Language Processing, such as mBERT \cite{bert}, XLM \cite{xlm}, ERNIE-m \cite{ernie}. Previous works on SER task to achieve multilingual training usually merge corpus and train together, to train on one corpus and test on others, or to train on one language and then fine-tune on other languages ~\cite{multilingual1}. Moreover, \cite{multitask} uses multi-task training by figuring out both the emotion class and the language type in a single model. \cite{multilingual2} utilizes attention mechanism to assign different weight to information extracted from different pieces of the input for each language, and then combines them in a weighted sum.

\section{Dataset and Features}
We evaluate our model on three languages: English IEMOCAP \cite{iemocap} dataset, German EmoDB \cite{emodb} dataset and French Cafe \cite{cafe} dataset. The three languages belong to two language families - Germanic(English, German) and Roman (French). For every dataset, we use the wavform audio file only. Table 1 shows the basic information for each dataset. For IEMOCAP, we reduce the emotion classes to four by merging Excited and Happy into a single category. For all the datasets, we split the test and dev set randomly with proportion 20:80, and split the dev set into validation set and training set randomly with proportion 20:80.

In our proposed architecture, we extract utterance-level acoustic features using pretrained models. Totally, five different features are extracted, including Allosaurus \cite{allosaurus}, MFCC, Wav2Vec \cite{wav2vec1}, GE2E \cite{ge2e}, and BYOL \cite{byol}. We use librosa\cite{librosa} to extract MFCC feature, and the other four features are extracted through pretrained models.

\begin{table*}
\centering
\resizebox{1.1\hsize}{!}{
\begin{tabular}{lcccc}
 \hline
           Dataset   & Language & Num of Speaker & Emotion Classes & Num of Utterances \\
    \hline
    \textbf{IEMOCAP} & English & 10 & Neutral, Happy, Anger, Sad & 5531 \\

    \textbf{EmoDB} & German & 10 & Neutral, Happy, Anger, Sad, Fear, Bored, Disgusted & 535 \\

    \textbf{Cafe} & French & 12 & Neutral, Happy, Anger, Sad, Fear, Surprise, Disgusted & 504 \\
    \hline
\end{tabular}
}
\label{tab:dataset}
\caption{Basic Information For Three Dataset}
\end{table*}

\section{The Proposed Approach}\label{method}
In this section, we first introduce our model architecture for speech emotion recognition in single language setting and propose a contrastive auxiliary loss. Next we introduce multi-domain training in detail, including multi-gating mechanism and neural architecture search module.

\subsection{Model Architecture}
We design a speech emotion recognition model (Figure 1) that uses raw waveforms as input and outputs predicted emotions. 

The first module of the model is the feature extractor. We firstly extract features from raw waveform via pretrained feature extractors.
For pretrained feature extractors that output 2d feature representations (e.g. Allosaurus, Wav2vec), 
we use a CNN + Bi-LSTM + self-attention network as encoding module to transform the original representations into 1d feature embeddings. More specifically, the 2-d feature representations generated by a certain pretrained feature extractor, denote as $z \in \mathbb{R}^{L \times d}$ ($L$ is the sequence length and $d$ is embedding size), is firstly used as an input to convolutional layer in order to extract n-gram features and thereby transform to $z' \in \mathbb{R}^{L\times d'}$, which is then used as input to the stacked Bi-LSTM layer for time aggregation. We treat the last hidden state of the stacked Bi-LSTM $\hat{z} \in R^k$ as the representation of the entire sequence, where $k$ is the hidden size of stacked Bi-LSTM. In order to further capture high-order information and integrate the entire output sequence of LSTM, we leverage the self-attention mechanism ~\cite{vaswani:transformer} by treating $\hat{z}$ as query, the outputs of LSTM at each time step as keys and values. The aggregated hidden state is computed as: 
\begin{equation}
\begin{aligned}
s(z') = Attention(\hat{z}\textbf{W}^Q, \textbf{H}^l\textbf{W}^K,\textbf{H}^l\textbf{W}^V)
\end{aligned}
\end{equation}

where the projection matrices  $\textbf{W}^Q,\textbf{W}^K,\textbf{W}^V $ are learnable parameters, and $\textbf{H}^l$ is the entire LSTM output sequence. The Attention function is \textit{Scaled Dot-Product Attention}:
\begin{equation}
\begin{aligned}
Attention(\textbf{Q},\textbf{K},\textbf{V}) = softmax(\frac{\textbf{Q}\textbf{K}^T}{\sqrt{k}})\textbf{V}
\end{aligned}
\end{equation}

where $\textbf{Q},\textbf{K},\textbf{V}$ are projected from $\hat{z}$,$\textbf{H}^l$,$\textbf{H}^l$ with $\textbf{W}^Q$, $\textbf{W}^K$, $\textbf{W}^V$ respectively. The output of self-attention layer is an aggregation of the entire LSTM output sequence that contains high-order information. 
We then concatenate feature representations $s(z')$ with 1-d feature representions $g$ (e.g. GE2E, BYOL) as the final audio representation $g \bigoplus s(z')$. The final audio representation is then used for computing similarity matrix for an contrastive auxiliary loss that attempts to draw representations of samples with same labels closer to each other, while pushing samples with different labels away from each other.

The second module of the model is emotion classification module.  Suppose there are $C$ emotion categories for a certain language, we apply a fully-connected network $h_\Phi$ that maps $g \bigoplus s(z')$ to the logits $c \in R^C$ . In this way, we obtain predictions for emotion classes in terms of probability distribution.

At inference time, we replaced softmax with the argmax operator and select the most probable emotion class label as the output.

\begin{figure}[h]
  \centering
  \includegraphics[width=1.0\linewidth]{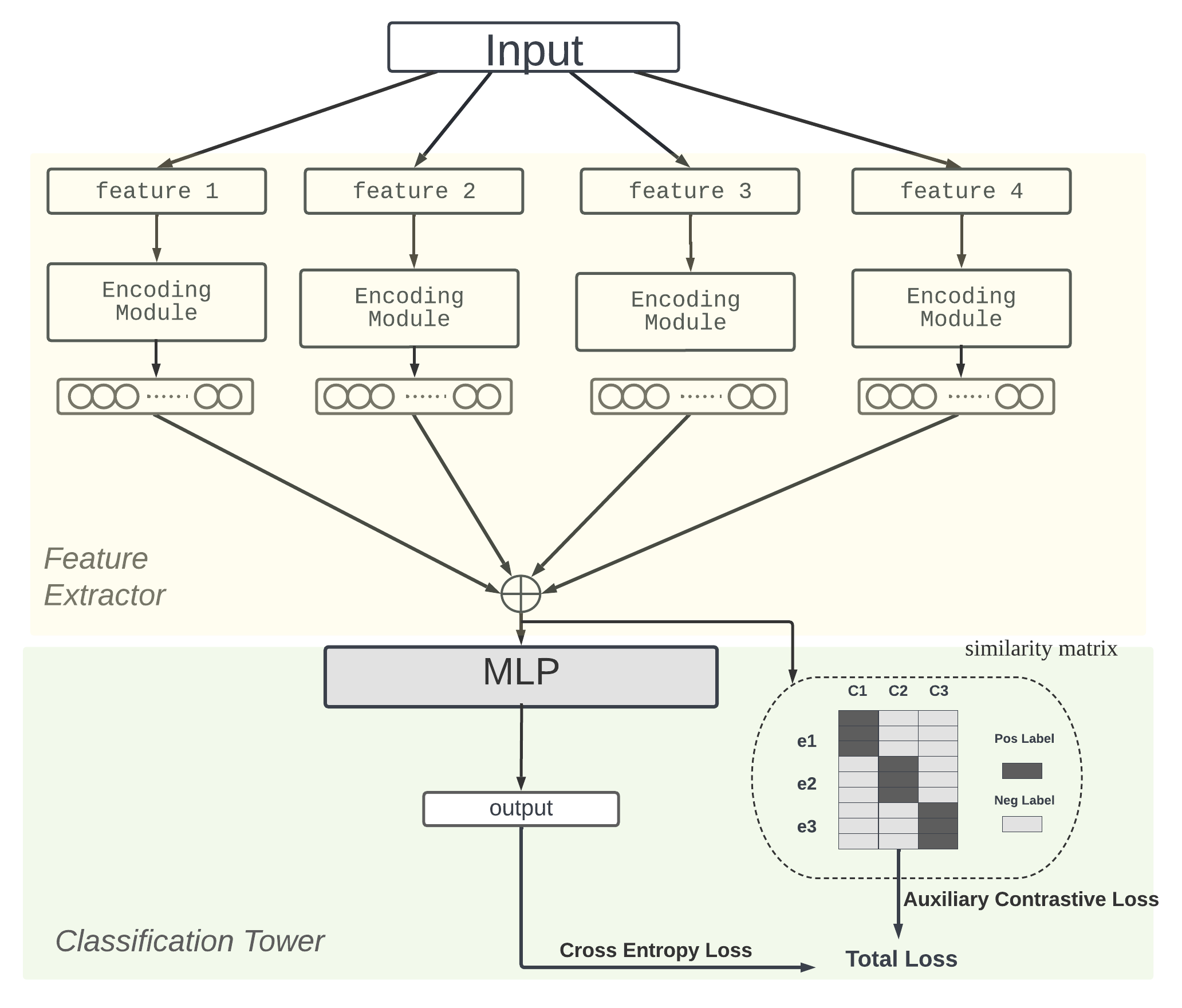}
  \caption{The structure of our model for single language SER. The feature extraction layer generates $M$ feature representations through pretrained models and encoding modules. At classification tower, the $M$ feature representations are firstly concatenated and then are send to feed forward layers. The concatenation of feature representations is also used to calculate the similarity matrix for auxiliary loss. The total loss is the combination of cross-entropy loss and the auxiliary loss.}
  \label{fig:singlemod}
\end{figure}

\subsection{Contrastive Auxiliary Loss}
Contrastive learning is a self-supervised learning technique that encourages augmentations of the same sample to have more similar representations than the augmentations of different samples. The idea of contrastive learning can also be transplanted into supervised-learning settings. Inspired by the contrastive training method of GE2E \cite{ge2e}, we add a contrastive auxiliary loss in order to learn a better feature representation. The auxillary loss encourages representation of wav samples of a certain emotion to be closer to its emotion centroid and away from  the centroid of other emotions.

More specifically, for each batch, we constructed a $N \times M \times d$ matrix, where $N$ is the number of emotions that appear in that batch and $M$ is the number of samples per emotion. In order to guarantee the robustness and stability of our model, we only retained the emotions with more than five samples in each batch. Each feature vector for the $i$-th sample of emotion $j$, denoted as $x_{ji} \in \mathbb{R}^d$, is the final representation that serves as input to classification tower with L2-normalization.

Then we constructed a similarity matrix $S \in \mathbb{R}^{(N\times M) \times N} $, where $S_{ji,k}$ is the affine transformation of cosine similarity between embedding vector $x_{ji}$ to the $k$th emotion centroids $c_k (1 \leq j,k \leq N, 1 \leq i \leq M)$:

\begin{equation}
\begin{aligned}
S_{ji,k} = w \cdot cos(x_{ji},c_k)+b
\end{aligned}
\end{equation}
where $w$ and $b$ are learnable parameters, and the centroid of emotion $k$ is calculated as:

\begin{equation}
\begin{aligned}
c_k = \frac{1}{M}\sum_{i=1}^M x_{ki}
\end{aligned}
\end{equation}

The loss of each embedding vector $x_{ji}$ is computed with cross-entropy loss in order to push each embedding vector closer to its emotion centroid, and pull it away from all other centroids:
\begin{equation}
\begin{aligned}
L(x_{ji}) = -log(\frac{exp(S_{ji,j})}{\sum_{k=1}^N exp(S_{ji,k})})
\end{aligned}
\end{equation}

We add this contrastive loss as an auxiliary loss to our model, in addition to the main cross-entropy classification loss. Therefore, the total loss is:
\begin{equation}
\begin{aligned}
L_{tot} = L_{CE}+\alpha \cdot \sum_{i,j} L(x_{ji})
\end{aligned}
\end{equation}

where $\alpha$ is the combine ratio of the auxiliary loss.


\subsection{Multilingual Training}
The previous sections illustrate our speech emotion recognition model for single language. However, using separate models for different languages makes it hard for model deployment and maintenance. Therefore, it is necessary to  apply  multilingual training to different languages in a single model in order to reduce model storage space and maintenance cost.

The most basic model is a share-bottom model ~\cite{mtl}. The concept is to build a shared bottom block to extract feature representations from data and then send the representation to classification towers that are specific for each domain. It is commonly used for multi-task and multi-domain learning because of its simplicity \cite{fastrcnn} \cite{fastrcnn2}.

In speech emotion classification scenario, suppose that there are $K$ languages, and the model consists of $M$ shared feature extractors, each feature extractor is represented as function $f_i$. There are $K$ classification towers $h_k$ for each domain respectively. Then for domain $k$, the model can be formulated as,$y_k = h_k(\sum_{i=1}^M f_i(x))$

The drawback of the shared bottom model is that the inherent conflicts caused by domain differences can harm the predictions of some domain, particularly when model parameters are extensively shared among all domains.

\subsubsection{Multi-gating Mechanism}

In order to solve the negative transfer problem in shared-bottom model especially when multiple domains have very different data distributions, we propose a multi-gating mechanism model that assign different weights to each feature extractor for different domains. The most similar work is Multi-gate Mixture of Experts (MMoE) \cite{mmoe} that shares the expert sub-models for different tasks while training a gating network to optimize each task. Different from MMoE, we apply multiple gating mechanism in multi-domain settings where data from different domains are differently distributed, while MMoE is used in multi-task settings where i.i.d distributed data is used to to predict different tasks.

\begin{figure}[h]
  \centering
  \includegraphics[width=1.2\linewidth]{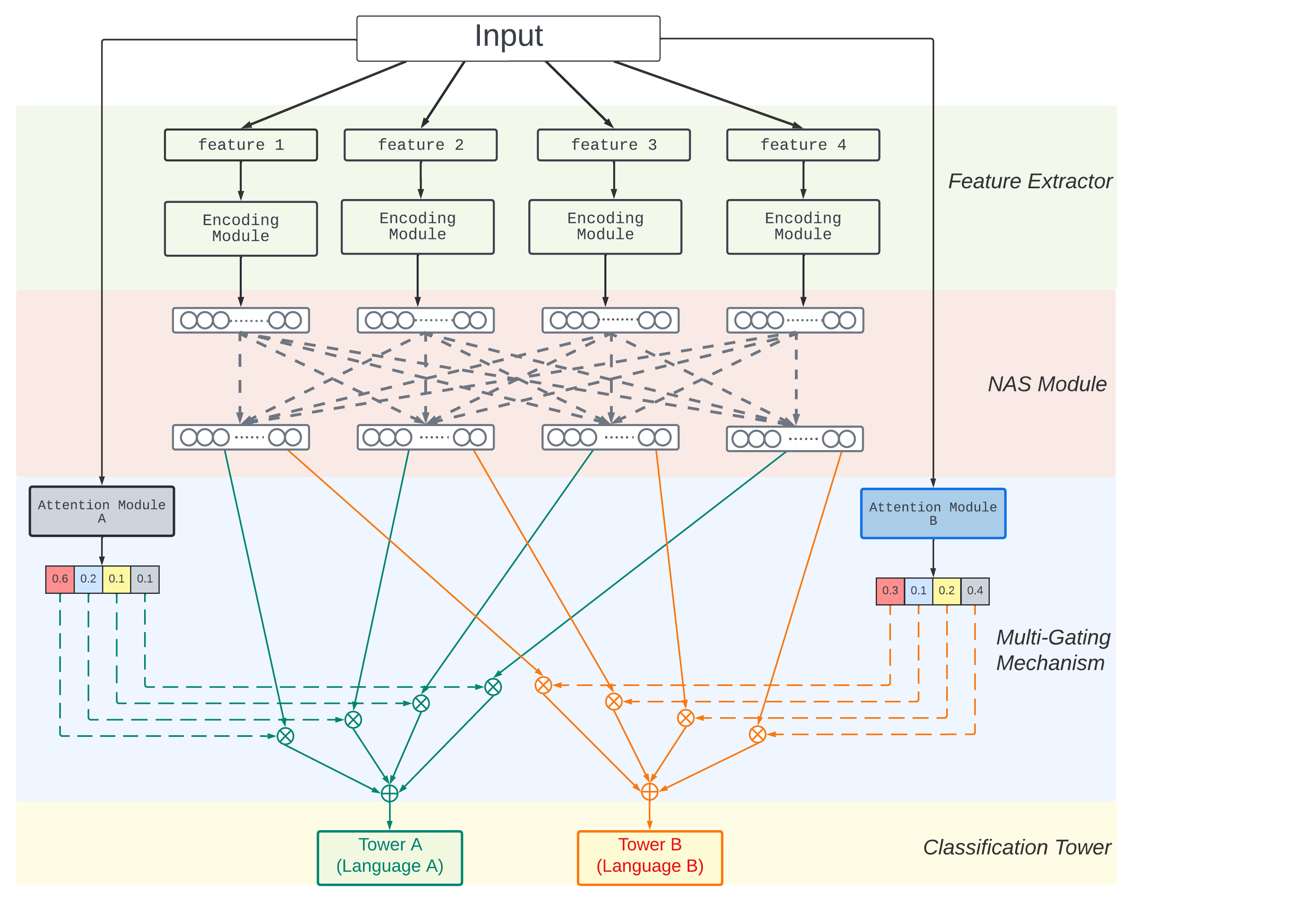}
  \caption{The structure of our model for multi-domain training. The feature extraction module is the same as Figure \ref{fig:singlemod}, the NAS module searchs for optimal neural architecture for each domain, the multi-gating mechanism learns different attention weights for different domains, and the classification tower are specific for each domain}
  \label{fig:mmoemod}
\end{figure}

In speech emotion recognition scenario, we treat each language as a specific domain since the data distribution of each language are different from others. The model firstly uses $M$ different feature extractors (e.g. MFCC, Allosaurus, WAV2VEC, GE2E, BYOL) to capture audio features from raw audio input, and then utilize domain-specific gating network to assign attention weights for the $M$ feature representation for each domain.  More specifically, the gating networks are linear transformations of the input with a softmax layer:
\begin{equation}
\begin{aligned}
g(x) = softmax(W_kx),
\end{aligned}
\end{equation}
where $W_k \in R^{M\times d}$is a trainable matrix, and $d$ is the dimension of $x$. We used GE2E embedding as attention selector $x$, since the GE2E feature is proved to be a good representation for raw audio. Therefore, the output of domain $k$ is $y_k = h_k(\sum_{i=1}^M g_i(x)f_i(x))$.

Since each gate is assigned exclusively to one domain, and the parameter of multiple gates are not shared across different domains, our model can capture domain-specific information for each different language and alleviate the domain adaption problem \cite{da1} \cite{da2} that is prevalent in multi-domain learning.

\subsubsection{Neural Architecture Search Module}
Neural Architecture Search (NAS)\cite{nas} is a subfield of AutoML and is concerned with automatically designing optimal neural network architectures by gradient descent\cite{darts}, evolution strategy \cite{ea} or reinforcement learning \cite{rl}. However, NAS is seldom used in the field of multi-domain learning.

Inspired by \cite{snr}, we let the model to automatically learn the connection between sub-networks for each domain in order to achieve more flexibility. In this way, the model not only assigns different parameters, but also assigns different neural architecture for different domains, and thus provides more flexibility. 

Suppose there are $M$ low-level feature embeddings
 $\mathbf{v_{1}},\mathbf{v_{2}}, \dots,\mathbf{v_{M}}$. A Neural Architecture Search module on top of the $M$ feature embeddings transforms the low-level feature embeddings into higher-level representations using transition matrices and Bernoulli variables:

\begin{equation}
\begin{aligned}
\mathbf{u_{i}} = \sum_{j = 1}^{M} \xi_{ij} \mathbf{W_{ij}} \mathbf{v_{j}}
\end{aligned}
\end{equation}

where $\mathbf{u_{i}} (1 \leq i \leq M)$ is the higher-level representation, $\mathbf{W_{ij}}$ is a transition matrix from the i-th low-level feature embedding to j-th higher-level feature embedding, and $\xi_{ij}$ is a Bernoulli variable that controls the connectivity. $\xi_{ij}$ equals to zero indicates the absence of connection between the i-th embedding and the j-th embedding, therefore $\xi_{ij}$ serves as network architecture selector that introduces connection sparsity. Therefore, it is possible to search for the neural architecture that works best for each domain. 

Like Neural Architecture Search, we also learn the architecture and model parameters together. Since Bernoulli variable $\xi_{ij}$ is not differentiable, we use the hard concrete distribution proposed in \cite{bernoulli} to smooth a Bernoulli distribution so that we can calculate the gradients directly:

\begin{equation}
\begin{aligned}
u \sim U(0, 1), s&=Sigmoid(log(u)-log(1-u)+log(\kappa))/\beta,\\
\bar{s} &= s(\delta-\gamma) + \gamma,\xi = min(1, max(0,\bar{s})
\end{aligned}
\end{equation}

where $u$ is a uniform distributed variable, $\beta, \gamma, \delta$ are hyper-parameters and $\kappa$ is a learnable parameter. 

\section{Experiments}

\subsection{Experimental Settings}

\textbf{Implementation Details.}
All the three languages share the same low-layer encoding modules. Allosaurus and WAV2VEC features utilize CNN layer with 64 kernels of size $3 \times 3$ and $5 \times 5$, followed by a LSTM layer with hidden size 128, and self-attention layer with hidden size 256. MFCC feature uses CNN layer with 32 kernels, and a 2-layer LSTM with hidden size 64 instead. For the NAS module, we let $\beta = 0.9, \gamma = -0.1, \delta = 2$. 

The classification tower of each language is allowed to have different hyperparameters in other to fit better to specific data distribution for each language dataset. The unit number of feed forward layers for English, German and French are 256, 512, 512 respectively. For English, we use Mish ~\cite{mish} and GeLU ~\cite{gelu} activation function to add more nonlinear transformation. For French and German, we use Tanh and ReLU as activation function. The dropout rate for each layer is set to be 0.1 in order to alleviate over-fitting problems. The weight of contrastive auxilliary loss $\alpha$ is set to be 0.1, 0.01, and 0.015 for English, German and French respectively. 

We use AdamW with learning rate $lr = 0.001$, $\beta_1=0.9$, $\beta_2 = 0.999$, $\epsilon=$1e-08, weight\_decay=1e-05 as optimizer for all languages. The GE2E module is finetuned with learning rate $lr*0.01$. We train the model for 20 epoch for all languages. 

\textbf{Metrics.}
Weighted Accuracy (WA) is a mean accuracy over different emotion classes with weights proportional to the number of samples in each class. Unweighted accuracy (UA) is the average accuracy of different classes. In our experiments, each configuration is tested 5 times and we report the mean result.

\subsection{Results}

\subsubsection{Single Domain results}
In Table 2, we compare our single domain model with previous works. Our models outperforms the listed works which are experimented on the same dataset for German and French. We also outperform the state-of-the-art results for the French dataset by a large margin, as both the unweighted and weighted accuracies are at least 10\% higher than those found in other literature. However, We did not compare results of the English dataset, because our work solely depends on auditory data, while most speech emotion recognition tasks with state-of-the-art results on IEMOCAP dataset utilize transcript \cite{lian2020context} \cite{sermtl}, which would greatly improve SER accuracy. Therefore, our work is not comparable with the state-of-the-art results.
\begin{table*}
\centering
\resizebox{1\hsize}{!}{
\begin{tabular}{lcccc}
\hline
    Language                    &       Method                                              &   Description                             &   UA      &   WA      \\
    \hline
    \multirow{4}{*}{German}     &   Rudd et al.\cite{rudd2022leveraged}                     &   CNN-based network + MLP                 &   0.928   &           \\
                                &   Meng et al.\cite{meng2019speech}                        &   ADRNN using 3D log-mel spectrograms     &   0.850   &           \\
                                &   Zhao et al.\cite{zhao2019speech}                        &   1D \& 2D CNN LSTM networks               &   0.923   &           \\
                                &   Ours                                                    &   See section \ref{method}                &   \textbf{0.957}   &   \textbf{0.954}   \\
    \hline
    \multirow{4}{*}{French}     &   Ng et al.\cite{ng2021investigation}                     &   Capsule network with Cauchy–Schwarz loss&   0.458   &           \\
                                &   Keesing et al.\cite{keesing2021acoustic}                &   SVM-R with wav2vec                      &           &   0.763   \\
                                &   Scheidwasser-Clow et al.\cite{scheidwasser2022serab}    &   BYOL-S                                  &   0.764   &           \\
                                &   Ours                                                    &   See section \ref{method}                &   \textbf{0.871}   &   \textbf{0.851}   \\
    \hline
    
\end{tabular}
}
\caption{Baseline methods from the literature, compared with the proposed method. Best results are \textbf{bold}.}
\label{table:aug}
\end{table*}

\subsubsection{Multi-Domain Results}
We compare our multi-domain training method with four baseline models. In order to make fair comparison between the models, we fix the task-specific classification towers for each language, and the shared part among different languages are illustrated as follows:
\begin{itemize}
  \item Base: For the baseline model, all languages are trained together. The union of labels(8 labels in total)is adopted. Both low-level feature extractors and classification tower are shared. 
  \item SB: Shared Bottom model. The parameter of low-level feature extractors are shared by all languages, and each language has its domain-specific classification tower.  
  \item OMoE. This model adds a single gating network for all languages, and the gating network parameter is shared by all languages.
  \item MMoE. This model treats each low-level feature extractors as an expert and uses different gating networks for different languages. The gating network parameters are specific for different languages.

\end{itemize}
As illustrated in table 3, the consistent improvement validates the efficacy of our method. The Base training strategy is the worst, this is probably because sharing both the low-level encoding modules and higher-level classification tower deprives the model of the ability to differentiate unique characteristic of each language dataset. By using Shared-Bottom model, we see a huge improvement on WA and UA due to the separation of classification tower for different languages.
OMoE adds a single gating mechanism that assigns weights for each feature representations, and therefore perform better than Shared-Bottom model. By separating the parameter for gating module, MMoE achieves better performance than OMoE. Our model achieves more flexibility than all the baseline models by introducing multi-gating mechanism to assign weights automatically and a NAS module to search for optimal neural architecture for each language.

\subsection{Analysis}

\subsubsection{Auxiliary Loss}
In this subsection, we study whether the contrastive auxiliary loss helps the SER task. 
Table 4 reports WA and UA of single-domain SER model with and without auxiliary loss, we notice that using auxiliary loss gives constant improvement. Recall that the contrastive auxiliary loss encourages wav samples from the same emotion class to learn similar representations, while letting the samples from different emotion classes to learn different representations. The T-SNE visualization in Fig 3 validates the effect of auxiliary loss: with $\alpha = 0.01$, the representations of wav samples are more compact (i.e. higher intra-class similarity and lower inter-class similarity) than with $\alpha = 0$.

\begin{figure}[t]
\centering
\label{Fig.sub.1}
\includegraphics[width=3.5cm,height = 3.5cm]{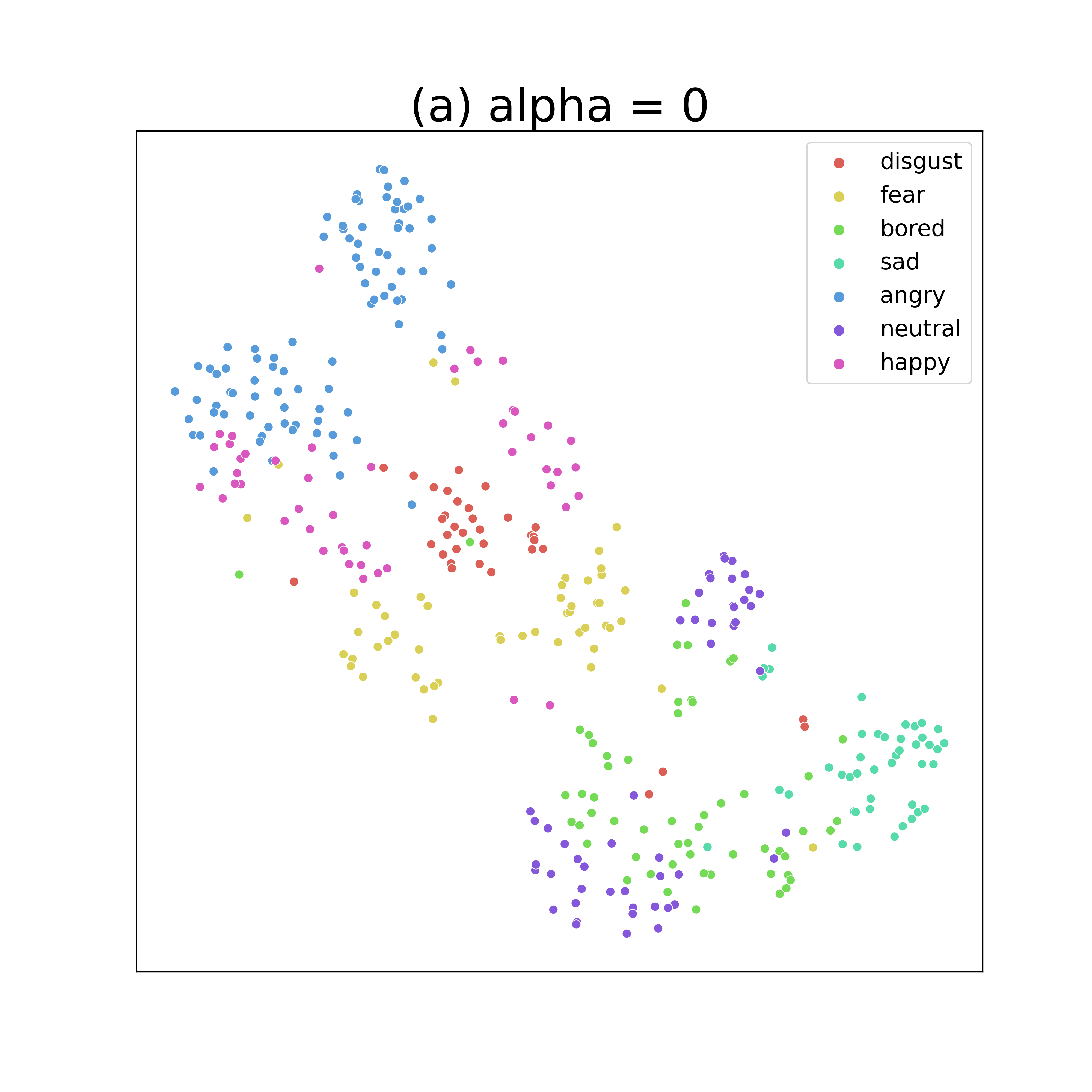}
\label{Fig.sub.2}
\includegraphics[width=3.5cm,height = 3.5cm]{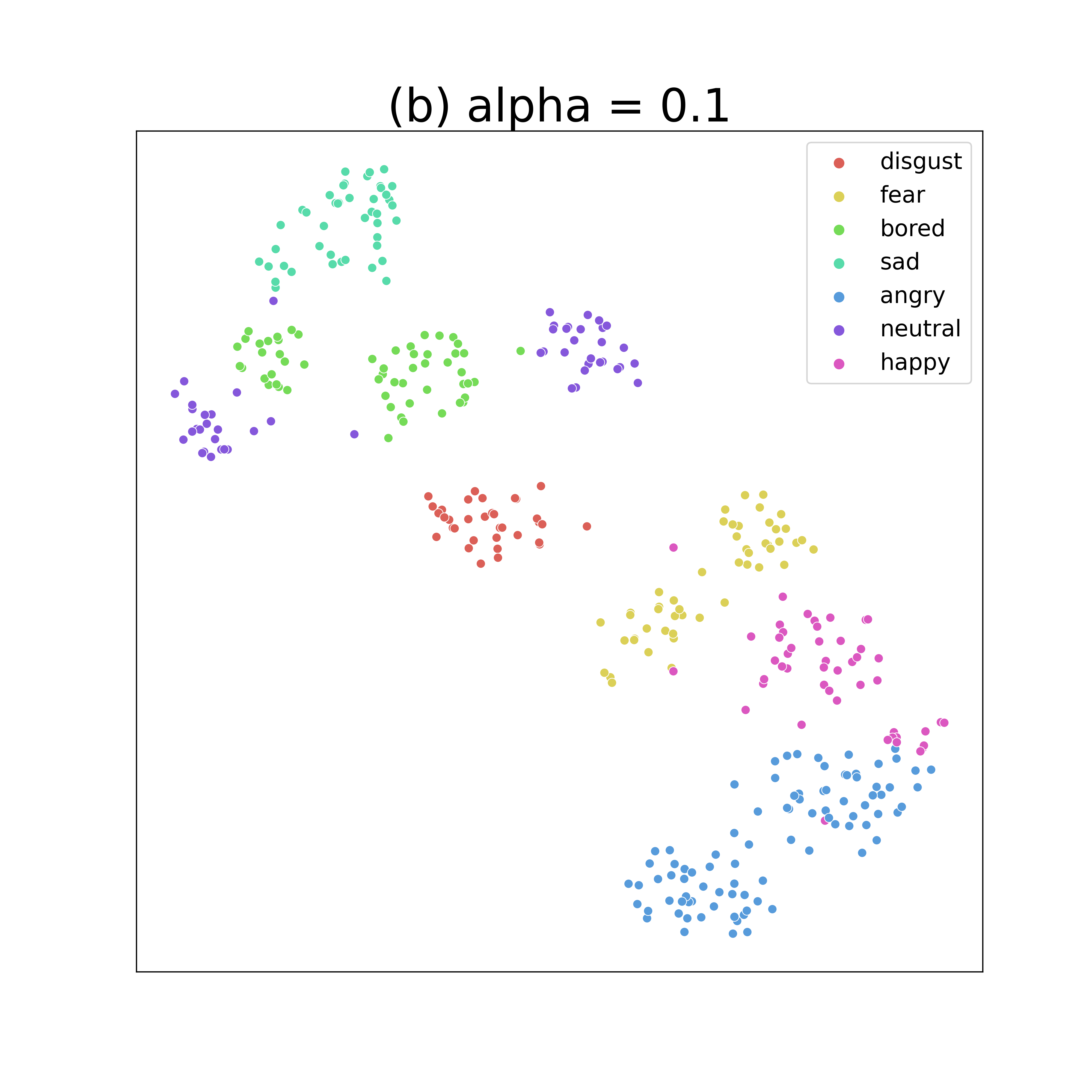}
\caption{T-SNE visualization of audio representations for 7 emotion categories with (a) $\alpha = 0$ and (b) $\alpha = 0.1$ for German Dataset.}
\label{1}
\end{figure}

\subsubsection{Gating weight distribution}
In order to understand how multi-gating mechanism helps multi-domain learning, we plot the average gating weight in the softmax gating network for each domain on each feature representation, as shown in Fig. 4. In this part, we use BYOL feature as selector $x$ in order to get the gating weight for the other four features according to formula (7). Note that it is feasible to use any feature as selector, but we only report the result when using BYOL feature as selector for simplicity. 
We see that the multi-gating networks learns attention weights for each domain automatically by assigning larger weights to features with better performance in a certain domain. 

\subsubsection{Gating weight distribution}
In order to understand how multi-gating mechanism helps multi-domain learning, we plot the average gating weight in the softmax gating network for each domain on each feature representation, as shown in Fig. 4. In this part, we use BYOL feature as selector $x$ in order to get the gating weight for the other four features according to formula (7). Note that it is feasible to use any feature as selector, but we only report the result when using BYOL feature as selector for simplicity. 
We see that the multi-gating networks learns attention weights for each domain automatically by assigning larger weights to features with better performance in a certain domain. 

\begin{figure}[h]
  \centering
  \includegraphics[width=1\linewidth]{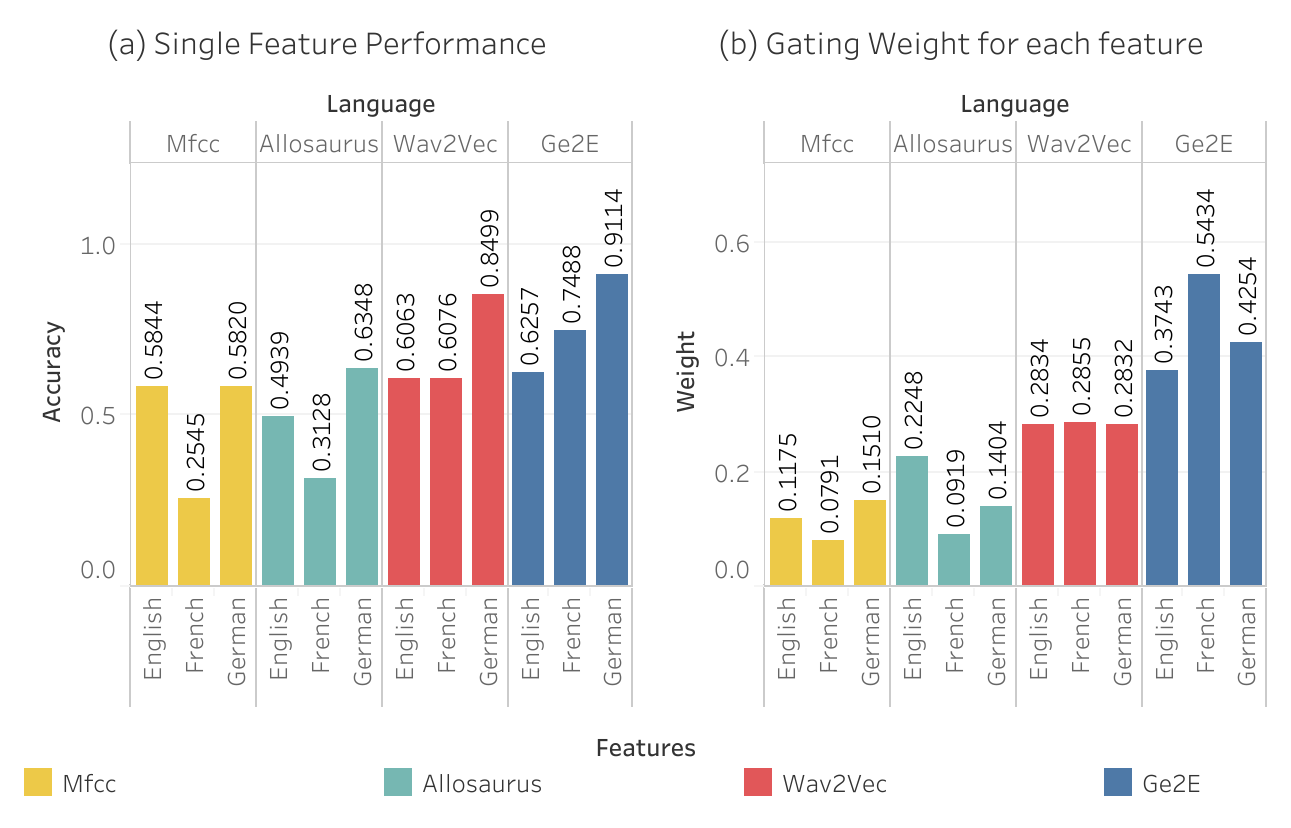}
  \caption{The comparison between (a) single feature performance and (b) gating weight assigned to each feature. Multi-gating mechanism assigns larger weights to feature with better performance (i.e. GE2E, WAV2VEC), and smaller weights to feature with worse performance (i.e. Allosaurus, MFCC).}
  \label{fig:feaimp}
\end{figure}

\section{CONCLUSION}

In this paper, we firstly propose a single-domain model for single language SER task, which leverages different pretrained speech models as the feature extractor backbone, and utilizes a contrastive auxiliary loss to learn better feature representations.  Secondly, we build a multi-domain model to perform SER tasks on different languages simultaneously. The multi-domain model uses a multi-gating mechanism and a neural architecture search module to alleviate negative transfer among different domains. 

To evaluate the effectiveness of our proposed method, we conduct experiments on English, German and French SER dataset.  Experimental results demonstrate that our method achieves new state-of-the-art accuracy on the French and German datasets, the two low-resourced dataset in our experiment. The experiment also shows that the contrastive auxiliary loss enables the model to learn more compact feature representations for each emotion class. Moreover, the experiment demonstrates that our multi-domain model has least negative transfer problem compared to other existing methods.

\begin{table}
\centering
\resizebox{1\hsize}{!}{
\begin{tabular}{lccc}
\hline
& English &  German & French  \\
\hline
    Base & 0.3333/0.6459 & 0.6403/0.8586 & 0.6860/0.7646 \\
    SB & 0.6507/0.6570 & 0.8628/0.8702 & 0.8538/0.8410  \\
    OMoE & 0.6723/0.6722 & 0.9398/0.9443 & 0.8580/0.8417  \\
    MMoE & 0.6883/0.6815 & 0.9419/0.9443
    & 0.8638/0.8415 \\
    Ours & \textbf{0.7045/0.7126} & \textbf{0.9578/0.9536} & \textbf{0.8713/0.8514} \\
    \hline
\end{tabular}
}
\caption{Quantitative evaluation of several multi-domain training methods, recorded in format UA/WA. Best results are \textbf{bold}. }
\end{table}

\begin{table}
\centering
\resizebox{1\hsize}{!}{
\begin{tabular}{lccc}
\hline
& English &  German & French  \\
\hline
with aux loss & \textbf{0.7434/0.7331} & \textbf{0.9556/0.9536} & \textbf{0.9070/0.8910} \\
    w/o aux loss & 0.7213/0.7219 & 0.9493/0.9536 & 0.8935/0.8712 \\
    \hline
\end{tabular}
}
\caption{Quantitative ablation of the auxiliary loss in  three datasets, recorded in format UA/WA}
\end{table}

\bibliography{acl2020}
\bibliographystyle{acl_natbib}

\end{document}